\documentclass[preprint,12pt]{elsarticle}
\usepackage{amssymb}
\usepackage{amsmath}
\usepackage{natbib}
\usepackage[margin=1in]{geometry}
\usepackage{algorithm}
\usepackage{algorithmic}
\usepackage{booktabs}
\usepackage{multirow}
\usepackage{caption} 
\usepackage{hyperref}

\begin{document}

\begin{frontmatter}

\title{BO-SA-PINNs: Self-adaptive physics-informed neural networks based on Bayesian optimization for automatically designing PDE solvers}

\author[lab1]{Rui Zhang} 
\ead{uestczr2021@163.com}
\author[lab1]{Liang Li\corref{cor1}} 
\ead{plum\_liliang@uestc.edu.cn, plum.liliang@gmail.com}
\author[lab2]{St\'ephane Lanteri} 
\ead{stephane.lanteri@inria.fr}
\author[lab1]{Hao Kang} 
\ead{kanghao2022@gmail.com}
\author[lab1]{Jiaqi Li} 
\ead{15079492558@163.com}

\address[lab1]{School of Mathematical Sciences, University of
               Electronic Science and Technology of China, 611731,
               Chengdu, P.R. China}

\address[lab2]{Universit\'e C\^ote d'Azur, Inria, CNRS, LJAD, Sophia Antipolis, France}

\cortext[cor1]{Corresponding author}

\begin{abstract}
Physics-informed neural networks (PINNs) is becoming a popular alternative method for solving partial differential equations (PDEs). However, they require dedicated manual modifications to the hyperparameters of the network, the sampling methods and loss function weights for different PDEs, which reduces the efficiency of the solvers. In this paper, we propose a general multi-stage framework, \emph{i.e.} BO-SA-PINNs to alleviate this issue. In the first stage, Bayesian optimization (BO) is used to select hyperparameters for the training process, and based on the results of the pre-training, the network architecture, learning rate, sampling points distribution and loss function weights suitable for the PDEs are automatically determined. The proposed hyperparameters search space based on experimental results can enhance the efficiency of BO in identifying optimal hyperparameters. After selecting the appropriate hyperparameters, we incorporate a global self-adaptive (SA) mechanism the second stage. Using the pre-trained model and loss information in the second-stage training, the exponential moving average (EMA) method is employed to optimize the loss function weights, and residual-based adaptive refinement with distribution (RAR-D) is used to optimize the sampling points distribution. In the third stage, L-BFGS is used for stable training. In addition, we introduce a new activation function that enables BO-SA-PINNs to achieve higher accuracy. In numerical experiments, we conduct comparative and ablation experiments to verify the performance of the model on Helmholtz, Maxwell, Burgers and high-dimensional Poisson equations. The comparative experiment results show that our model can achieve higher accuracy and fewer iterations in test cases, and the ablation experiments demonstrate the positive impact of every improvement.
\end{abstract}

\begin{highlights}
\item We propose a general multi-stage framework—BO-SA-PINNs—which uses BO (based on our proposed hyperparameters search space) to automatically determine the optimal network architecture, learning rate, sampling points distribution and loss function weights for the PDE solvers based on pre-training.
\item We introduce a new global self-adaptive mechanism. We use EMA to optimize the loss function weights and employ RAR-D to optimize the distribution of sampling points based on the pre-trained model and the loss information of the second stage;
\item We propose a new activation function suitable for PINNs;
\item We have verified the effectiveness of BO-SA-PINNs on 2D Helmholtz equation, 2D Maxwell equation in heterogeneous media, 1D viscous Burgers equation and high-dimensional Poisson equation, achieving a lower L2 relative error and fewer iterations compared to existing methods.
\end{highlights}


\end{frontmatter}



\section{Introduction}
PDEs play an important role in mathematical modeling \cite{evans2022partial} in fields such as electromagnetics, acoustics, and fluid mechanics. The core idea is to represent the process through a function that describes its behavior in space and time \cite{grossmann2024can}. In many practical cases, due to their complexity, PDEs are difficult to obtain analytical solutions, so numerical solutions are needed. One common strategy for solving PDEs is discretization, in which the continuous problem is transformed into a system of algebraic equations to solve. Common methods include the finite element method \cite{nikishkov2004introduction}, the finite difference method, and the spectral element method. However, traditional numerical methods struggle to solve high-dimensional, nonlinear, and non-smooth PDEs \cite{guo2020solving}.

With the booming development of scientific machine learning \cite{brunton2022data}, PINNs \cite{raissi2019physics} are proposed in 2019, which use neural networks with physical prior information to solve PDEs. PINNs can handle complex problem domains and boundary conditions, and have the potential to solve high-dimensional problems \cite{meng2022physics}, but their accuracy is unexpectedly low when solving stiff PDEs or handling complex problem definitions \cite{karniadakis2021physics}. In addition, in PINNs training process, the network architecture, loss function weights, sampling points distribution and activation function have a great influence on the accuracy \cite{wang2023expert}. If the hyperparameters are not well selected, overfitting or underfitting is likely to occur. Many existing studies select hyperparameters through extensive experiments, empiricism, or random search when PINNs lack strict theoretical convergence guarantees \cite{wang2022and}, but in fact they do not maximize the fitting ability of the network. There are also a few studies that use automatic machine learning to select the hyperparameters of PINNs. For example, a study proposed using Bayesian optimization to select neural network hyperparameters suitable for the Helmholtz equation, but ignored the sampling points distribution and loss function weights \cite{escapil2023hyper}. There are also studies using neural architecture search \cite{wang2022auto} and genetic algorithms \cite{le4590874hyperparameter} to select hyperparameters. For the influence of activation functions \cite{dung2023choice}, some studies have proposed self-adaptive activation functions to mitigate these challenges and improve performance \cite{wang2023learning}.

Additionally, PINNs have been improved through integration with other technologies. For example, some methods combined PINNs with domain decomposition techniques \cite{moseley2023finite}\cite{dolean2024multilevel}\cite{si2025initialization}, using multiple sub-networks to capture specific regional features, thereby improving training speed and accuracy. Some research used meta-learning methods to pre-train PINNs to build lightweight meta-networks \cite{chen2024gpt}, enabling the fast and low-cost generation of numerical solutions for PDEs. Other studies have proposed improvements to optimization algorithms. PINNs commonly use the ADAM optimizer and the L-BFGS optimizer, but ADAM is prone to getting stuck in the local optimal solution \cite{wang2021understanding}. One study combined genetic algorithms with L-BFGS for optimization \cite{pratama2023solving}. Some papers also improve the performance of PINNs by changing the basic network types, such as RNN \cite{wu2022physics} and ResNet \cite{zhang2023constrained}.

The integration of self-adaptive mechanisms with PINNs has also become a research hotspot. In addition to the adaptive activation function, some introduced a series of adaptive PINN schemes, including non-adaptive weighting of the training loss function, adaptive resampling of sampled points, and time-adaptive methods \cite{wight2020solving}. At the same time, one article \cite{braga2021self} discussed improved self-adaptive weight loss, called SA-PINNs. This method uses a soft attention mechanism to adjust the loss function weights of every sampling point, enhancing the performance of PINNs in difficult regions when approximating the solution. However, computational cost required for existing self-adaptive PINNs is too high, and we plan to further reduce it.

This paper introduces BO-SA-PINNs, a novel multi-stage self-adaptive PINNs framework that addresses the issues mentioned earlier with higher accuracy and lower computational cost. We briefly summarize the innovations and contributions of this study as follows.
\begin{itemize}
\renewcommand{\labelitemi}{\scriptsize$\bullet$}
\item  BO-SA-PINNs use BO to automatically determine the optimal network architecture, learning rate, sampling points distribution and loss function weights for PDE solvers based on pre-training. Our proposed hyperparameter search space, derived from experimental findings, enhances the efficiency of BO in identifying optimal hyperparameters. 
\item A new global self-adaptive mechanism is proposed. We use EMA to optimize the loss function weights and employ RAR-D to optimize the distribution of sampling points based on the pre-trained model and the loss information in the second stage.
\item We propose a new activation function suitable for PINNs.
\item We have verified the feasibility of the method on various benchmarks including the 2D Helmholtz equation, the 2D Maxwell equation in heterogeneous media (a complex numbers problem), the 1D viscous Burgers equation and high-dimensional Poisson equation, achieving a lower L2 relative error and fewer number of iterations compared to existing methods.
\end{itemize}

The remainder of the paper is organized as follows. In Section \ref{methodology}, we provide a brief review of neural network, PINNs and BO. The proposed framework and detailed algorithms are introduced in Section \ref{BO-SA-PINNs}. Moreover, the results of the numerical experiments including comparative and ablation experiments are presented in Section \ref{result}. In Section \ref{conclusion}, we draw some conclusion remarks and provide future plans.

\section{Methodology}
\label{methodology}
\subsection{Neural network}
\subsubsection{Fully-connected neural network}

Fully-connected neural network (FCNN) is one of the fundamental architectures in deep learning and is widely used due to its ability to approximate any continuous function as stated by the universal approximation theorem \cite{hornik1989multilayer}. It consists of multiple layers of interconnected neurons, where every neuron in a layer is connected to all neurons in the subsequent layer. Formally, a FCNN with \(L\) layers can be described as a composition of affine transformations followed by nonlinear activation functions.

Let the input be denoted as \( x \in \mathbb{R}^{d} \). Each layer \( l \) of the network performs the transformation:
\begin{equation}
    h^{(l)} = \sigma(W^{(l)} h^{(l-1)} + b^{(l)}),
\end{equation}
where, \( h^{(l)} \) represents the activation values of layer \( l \), with \( h^{(0)} = x \) as the input, \( W^{(l)} \in \mathbb{R}^{d_l \times d_{l-1}} \) is the weight matrix for layer \( l \), \( b^{(l)} \in \mathbb{R}^{d_l} \) is the bias vector, \( \sigma(\cdot) \) is a activation function.

To update the network parameters \(\theta\), the gradient of loss function \(\mathcal{L}(\theta)\) is computed via backpropagation and optimized using gradient-based methods such as gradient descent:
\begin{equation}
    \theta^{t+1} = \theta^t - \eta \nabla_\theta \mathcal{L}(\theta),
\end{equation}
where \(\eta\) is the learning rate, $\theta^t$ represents the model parameters at iteration $t$, $\theta^{t+1}$ is the updated parameter after the current iteration.

\subsubsection{Residual network}
Residual network (ResNet) \cite{he2016deep} is an advanced architecture designed to ease the training of deep neural networks by mitigating the vanishing gradient problem. The key idea behind ResNet is the introduction of \textit{skip connections}, which enables the network to learn residual mappings relative to the input variations, rather than directly approximating the complete underlying transformations. This architecture effectively resolves gradient vanishing in deep networks through the establishment of gradient highways, significantly enhancing the trainability of deep models. A typical residual block is formulated as:
\begin{equation}
y = \mathcal{F}(x, {W_i}) + x
\end{equation}
where \(x\) is the input to the block, \(\mathcal{F}(x, W_i)\) represents the residual mapping to be learned, and the addition of \(x\) acts as a shortcut connection.

\subsection{Physics-informed neural networks}

Consider the PDE with initial-boundary value as an example: find $u(x,t)$ such that
\begin{align}
\mathcal{N}[u(x,t)] &= s(x,t), \quad x \in \Omega, \; t \in [0,T]\\
\mathcal{B}[u(x,t)] &= g(x,t), \quad x \in \partial\Omega, \; t \in [0,T]\\
\mathcal{I}[u(x,0)] &= h(x), \quad x \in \Omega.
\end{align}
where the domain $\Omega \subset \mathbb{R}^d$ is a close set and $t \in [0,T]$. The operators $\mathcal{N}$, $\mathcal{B}$ and $\mathcal{I}$ are spatial-temporal differential operators. The unknown true solution $u(x,t)$ is approximated by the output $\hat{u}(x,t;\theta)$ of a PINN with inputs $x$ and $t$ and network parameters $\theta$.

Then, we construct the loss function by incorporating the PDE residual, boundary conditions, initial conditions and numerical data of the solution as follows:

\begin{gather}
\mathcal{L}_R(\theta) = \frac{1}{N_R} \sum_{k=1}^{N_R} 
\Bigl\lvert \hat{u}\bigl(x_R^k, t_R^k;\theta\bigr)-s\bigl(x_R^k, t_R^k\bigr) \Bigr\rvert^2 \\
\mathcal{L}_B(\theta) = \frac{1}{N_B} \sum_{k=1}^{N_B} 
\Bigl\lvert \hat{u}\bigl(x_0^k, t_B^k;\theta\bigr) - g\bigl(x_0^k, t_B^k\bigr)\Bigr\rvert^2 \\
\mathcal{L}_I(\theta) = \frac{1}{N_I} \sum_{k=1}^{N_I} 
\Bigl\lvert \hat{u}\bigl(x_I^k, t_0^k;\theta\bigr) - h\bigl(x_I^k, t_0^k\bigr)\Bigr\rvert^2 \\
\mathcal{L}_D(\theta) = \frac{1}{2N_D} \sum_{k=1}^{N_D} 
\Bigl\lvert \hat{u}\bigl(x_D^k, t_D^k;\theta\bigr) - u_D^k \Bigr\rvert^2.
\end{gather}
where $\mathcal{L}_R(\theta)$, $\mathcal{L}_B(\theta)$, $\mathcal{L}_I(\theta)$ and $\mathcal{L}_D(\theta)$ are the PDE residual loss, boundary loss, initial loss of sampling points and MSE of numerical data, respectively. Moreover, $\{(x_R^k, t_R^k)\}_{k=1}^{N_R}$ is the set of $N_R$ residual training points in the domain $\Omega \times (0,T)$, $\{(x_B^k, t_B^k)\}_{k=1}^{N_B}$ is the set of $N_B$ boundary points, $\{(x_I^k, t_I^k)\}_{k=1}^{N_I}$ is the set of $N_I$ initial points and $\{(x_D^k, u_D^k)\}_{k=1}^{N_D}$ is the set of numerical data of the solution. The total loss function for the PINN is then defined as:
\begin{equation}
\mathcal{L}(\theta) = \omega_R \mathcal{L}_R(\theta)
+ \omega_B \mathcal{L}_B(\theta)
+ \omega_I \mathcal{L}_I(\theta)
+ \omega_D \mathcal{L}_D(\theta)
\end{equation}
where $\omega_R$, $\omega_B$, $\omega_I$ and $\omega_D$ are weights corresponding to the four losses. The value of these four weights will affect the learning effect of PINNs \cite{yu2022gradient}, but there is no exploration of initialization of loss weights in existing research as far as we know. 

The next step is to use the neural network to optimize the loss function $\mathcal{L}(\theta)$ and then get the final prediction solution. For some problems, the PDEs only have boundary values which can be handled similarly to the above.

\subsection{Bayesian optimization}
\label{subsec2}
BO \cite{frazier2018tutorial} is a class of machine-learning-based optimization methods focused on solving the problem:
\begin{equation}
\max_{h \in \mathcal{H}} f(h)
\end{equation}
where $\mathcal{H}\subset\mathbb{R}^{D}$ is the hyperparameter search space with $D\leq 10$ according to \cite{frazier2018tutorial}. Evaluating the objective function \( f(h) \) tends to be computationally expensive and \( f(h) \) is often treated as a black-box, meaning it lacks known convex or linear structure and derivatives are unavailable. The primary goal of BO is to identify the global optimum, rather than being confined to local optima.

The BO framework primarily consists of two key components: surrogate model and acquisition function. Surrogate model is used to model the objective function in $\mathcal{H}$. In other words, it is a learning model whose inputs are the observed function values $\bigl(h_i, f(h_i)\bigr)$ and it provides an estimate of $f(h)$ over the space $\mathcal{H}$ once trained. 

A popular surrogate model in BO is the Gaussian process regression(GPR) \cite{ru2020bayesian}. Given a set of observed data points \(\{(h_i, y_i)\}_{i=1}^{n}\) with \(y_i = f(h_i)\), the GPR provides predictions at a new input \(h\) with mean and variance given by:
\begin{align}
\mu(h) &= k(h, H)\Bigl[K(H, H) + \sigma_n^2 I\Bigr]^{-1} y \\
\sigma^2(h) &= k(h,h) - k(h, H)\Bigl[K(H, H) + \sigma_n^2 I\Bigr]^{-1} k(H, h)
\end{align}
where, \(H = [h_1, h_2, \dots, h_n]^\top\) denotes the matrix of observed inputs and \(y = [y_1, y_2, \dots, y_n]^\top\) is the corresponding vector of outputs. The function \(k(\cdot,\cdot)\) represents the kernel function, with \(K(H, H)\) being the associated kernel matrix. Additionally, \(\sigma_n^2\) stands for the observation noise variance, and \(I\) is the identity matrix.
After the initial surrogate model is constructed, the acquisition function can be derived to select the next evaluation point. This function is a heuristic evaluation function derived from the surrogate model. For example, the expected improvement (EI) is defined as follows. Let $f\left(h^+\right)$ denote the best function value observed so far. Then, the expected improvement function at the candidate point $h$ is defined as:
\begin{equation}
EI(h) = (\mu(h) - f(h^+)) \, \Phi\!\left(\frac{\mu(h) - f(h^+)}{\sigma(h)}\right) + \sigma(h) \, \phi\!\left(\frac{\mu(h) - f(h^+)}{\sigma(h)}\right)
\end{equation}
where, \(\mu(h)\) and \(\sigma(h)\) are the predictive mean and standard deviation at \(h\), \(\Phi(\cdot)\) and \(\phi(\cdot)\) are the cumulative distribution function and probability density function of the standard normal distribution, respectively.

\section{BO-SA-PINNs: specific implementation process}
\label{BO-SA-PINNs}
Our training process is divided into three stages. The first stage employs BO to determine hyperparameters of network, sampling point distribution and loss function weights. In the second stage, we use global self-adaptive mechanisms to refine sampling point distribution and loss function weights based on real-time training feedback. The third stage is stable optimization. Moreover, the first two stages utilize the ADAM optimizer, while the third stage employs the L-BFGS optimizer. In training process, ADAM uses momentum mechanism and adaptive learning rate to quickly complete loss reduction and preliminary exploration of parameter space at a low computational cost; while L-BFGS achieves superlinear convergence in the smooth region of the loss function by approximating the second-order curvature information of the Hessian matrix, and finally reaches a high-precision numerical solution under physical constraints. In this part, we take the PDE with initial-boundary conditions as an example, other situations can be solved similarly.

\subsection{Stage 1: select optimal hyperparameters based on BO}

\subsubsection{Hyperparameters search space}

There have been studies that apply BO to select hyperparameters in machine learning algorithms including neural networks \cite{snoek2012practical}, but for PINNs, the hyperparameters involved are not only related to neural networks. As mentioned above, the determination of PINNs hyperparameters is time-consuming, so we use BO to determine the hyperparameters based on pre-training. Taking the FCNN and ResNet as examples, we only need to determine the input and output of the network, the fundamental network architecture, problem domain, $\mathcal{L}_R$, $\mathcal{L}_B$ and $\mathcal{L}_I$ based on the PDEs characteristics. Next, we can execute the first stage. The optional hyperparameter search spaces are shown in Table \ref{hyperparameters}.

\begin{table}[htbp]
\caption{Hyperparameter Settings}
\centering
\begin{tabular}{p{4cm} p{7cm} p{3cm}}
\hline
\textbf{Type} & \textbf{Hyperparameter} & \textbf{Range}\\
\hline
\multirow{2}{*}{Network architecture} 
  & Hidden layers(FCNN) / Residual blocks(ResNet) 
  & [2,8] / [1,4]\\
  & Hidden neurons 
  & [20,80]\\
\hline
\multirow{1}{*}{Optimizer} 
  & Learning rate for ADAM
  & [0.001, 0.01]\\
\hline
\multirow{3}{*}{Sampling points} 
  & Domain sampling points 
  & [500, 5000]\\
  & Boundary condition sampling points 
  & [100, 3000]\\
  & Initial condition sampling points 
  & [100, 1200]\\
\hline
\multirow{3}{*}{Loss function} 
  & PDE residual loss weight 
  & [0.01, 0.10]\\
  & Boundary condition loss weight 
  & [0.01, 0.25]\\
  & Initial condition loss weight 
  & [0.01, 0.10]\\
\hline
\end{tabular}
\label{hyperparameters}
\end{table}

Since BO may fall into a local optimal solution and then affect the overall performance \cite{frazier2018tutorial}, the selection of the types and ranges of hyperparameters is very important in order to ensure the performance of BO. The hyperparameter search spaces shown in Table 1 are summarized from rough experiments. If you need to select more than 10 hyperparameters, we recommend you use REMBO \cite{wang2016bayesian} and HeSBO \cite{nayebi2019framework} because they can better guarantee BO performance in high-dimensional search space. If you have prior knowledge about the optimal hyperparameters, then we recommend that you use prior-BO \cite{souza2021bayesian} to explore the search spaces because it can save a lot of time and further explore new hyperparameter groups. If the PDEs are high-dimensional, we recommend increasing the range of sampling points appropriately.

\subsubsection{Determine hyperparameters based on pre-training}

The objective function of BO is defined as minimizing the L2 relative error:
\begin{equation}
\max_{x \in A} -f(x,t;h) = - \frac{\|\hat{u}(x,t;\theta,h) - u(x,t)\|_{2}}{\|u(x,t)\|_{2}}
\end{equation}
where $h$ is a set of hyperparameters selected by EI.

\begin{figure}[t]
\centering
\includegraphics[width=\linewidth]{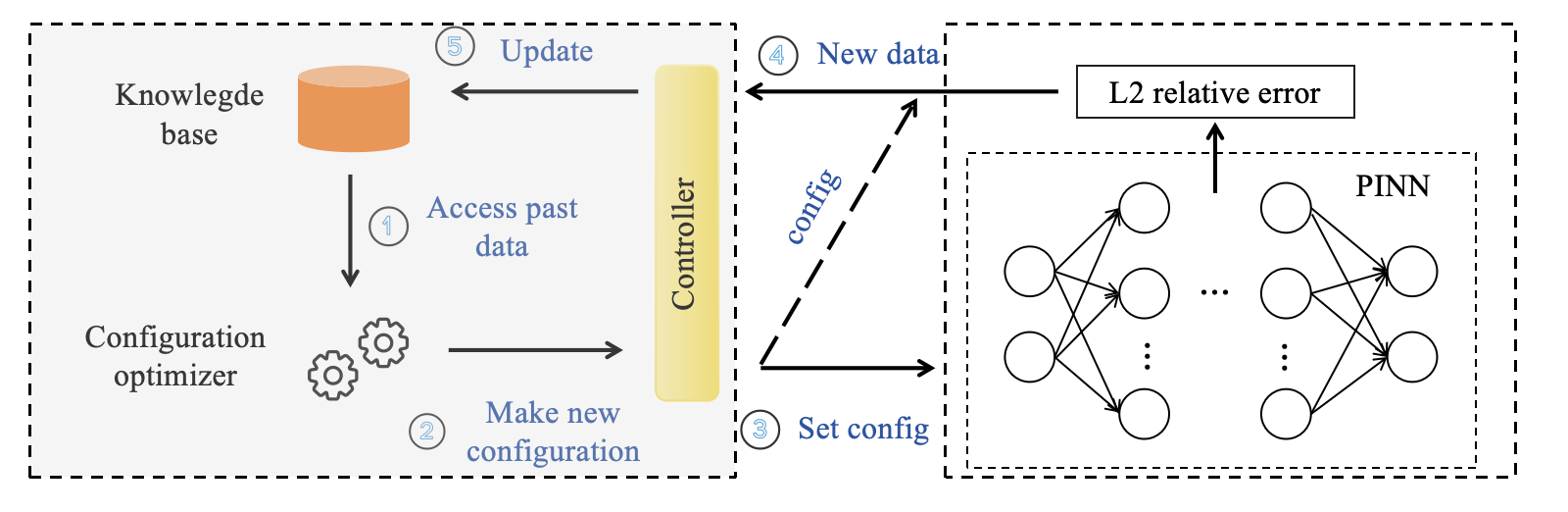}
\caption{Automated machine learning process}\label{fig:AMLP}
\end{figure}

Essentially, BO for selecting hyperparameters is a kind of automated machine learning process that simulates manual hyperparameter tuning while reducing the number of hyperparameter configurations to try, enabling the discovery of suitable hyperparameter configuration with lower computational cost. The tuning process is an iterative procedure, as illustrated in Figure \ref{fig:AMLP}. The knowledge base consists of configuration-performance tuples. The configuration optimizer selects new configurations, while the controller emulates manual experimentation. In the process, the configuration optimizer uses the current knowledge base to propose a configuration expected to improve performance. A PINN is then generated according to this configuration. After optimizing the PINN for fixed iterations, the L2 relative error is recorded and added as new data to the knowledge base, which subsequently updates based on the observations. This process repeats iteratively until the preset maximum number of iterations is reached.

In BO, we choose GRP as the surrogate model (\(f(x,t;h)\) can be proven to be continuous), EI as the acquisition function. The initial point is set to 10, and the optimization will run for 20 iterations to determine the optimal hyperparameters. In each evaluation, a specific group of hyperparameters is used to form a PINN to train for 500 iterations using the ADAM optimizer, and the L2 relative error is calculated. The hyperparameter group that yields the smallest L2 relative error will be selected for subsequent training. Additionally, the pre-trained model corresponding to the optimal hyperparameters after training 500 iterations will be saved as the initial model for subsequent training. The BO algorithm for selecting hyperparameters is outlined in Algorithm \ref{BO}.

\begin{algorithm}[htbp]
\small
\caption{BO for Selecting PINN Hyperparameters}
\begin{algorithmic}[1]
\STATE \textbf{Input:} $\mathcal{H}$: hyperparameter search space, $n_{\text{init}}$: initial sampling points in $\mathcal{H}$, $n_{\text{iter}}$: number of BO iterations
\STATE \textbf{Output:} $h_{\text{best}}$: optimal hyperparameters, $N(\cdot; \theta_{\text{best}}, h_{\text{best}})$: optimal pre-trained network

\STATE \textbf{Step 1: Initialize surrogate model}
\STATE Generate $n_{\text{init}}$ initial points $h_i$ in $\mathcal{H}$ using Latin Hypercube Sampling (LHS)
\FOR{$i = 1, \ldots, n_{\text{init}}$}
    \STATE Train $\mathcal{N}(x,t; \theta, h_i)$ using ADAM optimizer, record L2 relative error $f(h_i)$ after 500 iterations
\ENDFOR
\STATE Fit data $D_0 = \{(h_i, f(h_i))\}_{i=1}^{n_{\text{init}}}$ into a Gaussian process regression model

\STATE \textbf{Step 2: Update surrogate model based on acquisition function}
\FOR{$j = 1, \ldots, n_{\text{iter}}$}
    \STATE Obtain the acquisition function $a(h)$ defined by the current surrogate model
    \STATE Find new point $h_j$ maximizing expected improvement: $h_j = \arg\max_{h \in \mathcal{H}} a(h)$
    \STATE Train $\mathcal{N}(x,t; \theta, h_j)$ with ADAM optimizer for 500 iterations and record $f(h_j)$
    \STATE Update Gaussian process regression model with $D_j = D_{j-1} \cup \{(h_j, f(h_j))\}$
\ENDFOR

\STATE \textbf{Step 3: Return result}
\STATE $h_{\text{best}} = \arg\max_{h \in \mathcal{H}} f(h)$, $\theta_{\text{best}}$ is the network parameter corresponding to the best hyperparameter $h_{\text{best}}$
\STATE Return optimal hyperparameters $h_{\text{best}}$ and optimal network $N(\cdot; \theta_{\text{best}}, h_{\text{best}})$
\end{algorithmic}
\label{BO}
\end{algorithm}

\subsection{Stage 2: global self-adaptive mechanisms}
BO efficiently searches for optimal hyperparameters within a limited number of evaluations. However, due to issues such as insufficient evaluations, restrictive model assumptions, and susceptibility to local optima, BO cannot guarantee finding the global optimum. Heuristic self-adaptive mechanisms can help mitigate these limitations to some extent. The global self-adaptive mechanisms introduced in the second stage leverage loss information during training and residual error information from pre-trained model to further explore and adjust hyperparameters.

\subsubsection{EMA for adjusting loss function weights}
This method is designed to enable the optimal loss function weights to be heuristically updated according to the values of the residual loss, boundary loss or initial loss. For example, if the $\mathcal{L}_R$ is larger, we hope that the network can pay more attention to this part, so we can increase the proportion of $\omega_R$. Thus, this method allows the weights to dynamically adjust, with a bias towards the parts of the training that perform poorly.

Specifically, the weights for the residual, boundary and initial loss ($w_{R},w_{B},w_{I}$) are adaptively adjusted based on their recent historical values captured via EMA \cite{brown1959statistical}. Given the current iteration's $\mathcal{L}_R$, $\mathcal{L}_B$ and $\mathcal{L}_I$, respectively, we compute their moving averages $m_{R}$, $m_{B}$ and $m_{I}$. Using these averages, new provisional weights are computed proportionally. To avoid abrupt fluctuations and improve stability, these provisional weights are smoothly updated with the previous weights via another EMA controlled by a factor $\gamma \in (0,1)$. The algorithm of EMA for updating loss function weights is summarized in Algorithm \ref{EMA}.

\begin{algorithm}[htbp]
\caption{EMA for updating loss function weights}
\begin{algorithmic}[1]
\STATE \textbf{Input}: Initial values \( \omega_{R0}, \omega_{B0}, \omega_{I0}\) determined by BO, EMA decay parameter \( \beta \in (0, 1) \) for computing moving averages, EMA decay parameter \( \gamma \in (0, 1) \) for updating weights
\STATE \textbf{Output}: Updated weights \( \omega_{R}, \omega_{B},\omega_{I} \)
\FOR{0:1000:5000 iterations in training the optimal model from stage 1}
    \STATE Compute \( \mathcal{L}_R \), \( \mathcal{L}_B \) and \( \mathcal{L}_I \).
    \STATE Update moving averages and compute provisional weights:
    \[
    m_R^{(t)} = \beta \cdot m_R^{(t-1)} + (1 - \beta) \cdot \mathcal{L}_R^{(t)}, \quad m_B^{(t)} = \beta \cdot m_B^{(t-1)} + (1 - \beta) \cdot \mathcal{L}_B^{(t)}, \quad m_I^{(t)} = \beta \cdot m_I^{(t-1)} + (1 - \beta) \cdot \mathcal{L}_I^{(t)}
    \]
    \[
    \omega_R'^{(t)} = \frac{m_R^{(t)}}{m_R^{(t)} + m_B^{(t)}+m_I^{(t)}}, \quad \omega_B'^{(t)} = \frac{m_B^{(t)}}{m_R^{(t)} + m_B^{(t)}+m_I^{(t)}},\quad \omega_I'^{(t)} = \frac{m_I^{(t)}}{m_R^{(t)} + m_B^{(t)}+m_I^{(t)}}
    \]
    \STATE Update weights:
    \[
    \omega_R^{(t)} = \gamma \cdot \omega_R^{(t-1)} + (1 - \gamma) \cdot \omega_R'^{(t)}, \quad \omega_B^{(t)} = \gamma \cdot \omega_B^{(t-1)} + (1 - \gamma) \cdot \omega_B'^{(t)}, \quad \omega_I^{(t)} = \gamma \cdot \omega_I^{(t-1)} + (1 - \gamma) \cdot \omega_I'^{(t)}
    \]
    \STATE Clamp updated weights within predefined ranges to ensure numerical stability
\ENDFOR
\STATE \textbf{Return}: \( \omega_{R}, \omega_{B},\omega_{I} \)
\end{algorithmic}
\label{EMA}
\end{algorithm}

\subsubsection{RAR-D for adjusting sampling points distribution}
In the BO process, we obtain a pre-training model and the number of domain sample points. BO is a greedy algorithm that tends to seek globally optimal sample solution and employs the Hammersley sequence \cite{wong1997sampling} for sampling after determining the number of sampling points for each section. Consequently, high-error regions may not receive sufficient sampling points to learn local features, leading to inadequate training in those areas. Therefore, an adaptive sampling strategy can be introduced: by analyzing the residual error of the pre-training model, we dynamically add new training points in high-error regions, thereby improving the overall approximation accuracy of the network. 

However, relying solely on residual maximization to select new points may overlook other potential high-error regions or cause an excessive focus on a small subset of points. To address this, RAR-D \cite{wu2023comprehensive} proposes a hybrid approach that combines residual-driven refinement with random distribution sampling, aiming to balance the precise localization of high-error regions with comprehensive coverage of the problem domain. The core idea of RAR-D is to first construct a probability distribution based on the residual and then randomly sample new training points from that distribution. The algorithm is summarized in Algorithm \ref{RAR-D}.

\begin{algorithm}
\caption{RAR-D Algorithm}
\begin{algorithmic}[1]
\STATE \textbf{Input:} Optimal pre-trained model ${N}^*$, domain sampling points, $n_{\text{iter}}$
\STATE \textbf{Output:} New domain sampling points distribution
\FOR{$i = 1$ to $n_{\text{iter}}$}
    \STATE Randomly generate 1000 candidate points$\{x_j\}_{j=1,...,1000}$;
    \STATE Use the optimal pre-trained model to calculate PDE residuals of the candidate points $x_j$: $r(x_j) = |[\mathcal{N}^*(x_j)] - f(x_j)|^2$
    \STATE Normalize the PDE residuals at each point
    $\hat{r}(x_j) = \frac{r(x_j)}{\max \left( r(x_j) \right)}$
    \STATE Form a probability distribution function
    $P(x_j) = \frac{\hat{r}(x_j)}{\sum_{j=1}^{1000} \hat{r}(x_j)}$
    \STATE Use the constructed PDF to randomly sample 50 new training points based on the probability distribution: $x_{\text{new}} \sim P(x_j)$
    \STATE Add the new training points
\ENDFOR
\STATE \textbf{Return:} New domain sampling points distribution.
\end{algorithmic}
\label{RAR-D}
\end{algorithm}

\subsection{New activation function—TG activation function}
\label{subsec4}
\begin{figure}[t]
\centering
\includegraphics[width=0.55\linewidth]{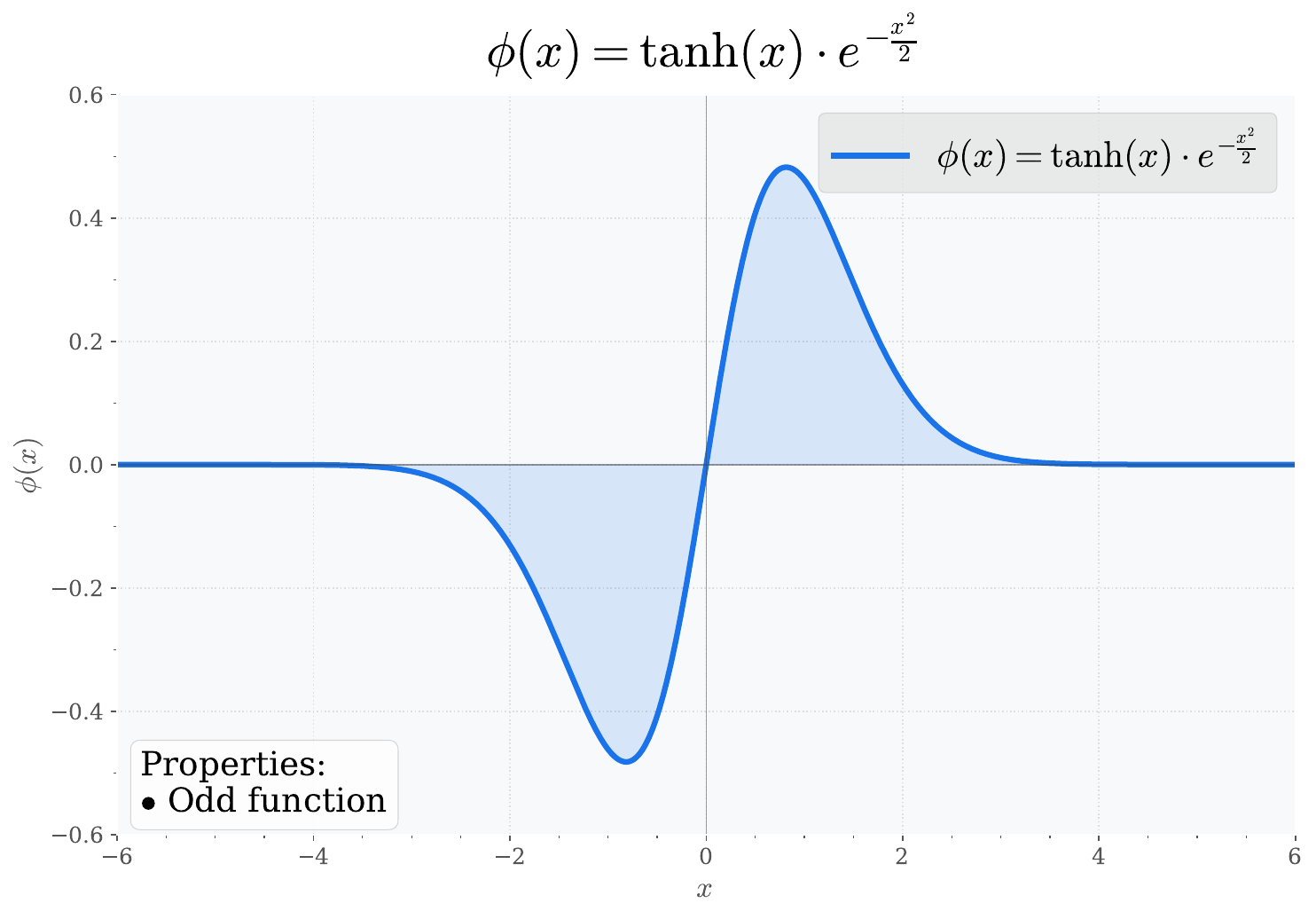}
\caption{TG activation function}\label{fig:TG}
\end{figure}

Currently, PINNs typically achieve the best accuracy using $\tanh(x)$ \cite{jagtap2020adaptive}. For problems with highly oscillatory solutions, $\sin(x)$ is sometimes considered. However, in practice, we have discovered a new activation function --- a $\tanh$ function with a Gaussian window --- that can further improve the accuracy of PINNs. The function is defined as:
\begin{equation}
\phi(x) = \tanh(x)\, e^{-\frac{x^2}{2}}.
\end{equation}
We call it the TG activation function shown in Figure \ref{fig:TG} and have proved its effectiveness in \ref{appA}. By combining the nonlinear mapping capability of $\tanh$ with the localized, smoothly decaying characteristic of the Gaussian function, this activation function provides PINNs with enhanced ability to capture local features. As a result, it is well suited for solving a wide variety of PDE problems, enabling stable training and improving accuracy .

\subsection{Whole process of BO-SA-PINNs}
Our BO-SA-PINNs only require the experimenter to determine the input and output of the network, the fundamental network architecture, problem domain, $\mathcal{L}_R$, $\mathcal{L}_B$ and $\mathcal{L}_I$ based on the PDE characteristics. Then our framework can automatically form a high-accuracy PINN. For the combination of ADAM and L-BFGS optimizer, this phased strategy of exploration followed by refinement significantly balances the training efficiency and the accuracy of the solution in solving complex PDEs through ADAM's global rough adjustment and L-BFGS's local refinement. The algorithm of BO-SA-PINNs is summarized in Algorithm \ref{whole process}.

\begin{algorithm}[htbp]
\caption{Complete Process of BO-SA-PINNs}
\begin{algorithmic}[1]
\STATE \textbf{Input}: Network input and output, problem domain, $\mathcal{L}_R$, $\mathcal{L}_B$ and $\mathcal{L}_I$
\STATE \textbf{Output}: Prediction solution $\hat{u}(x,t;\theta)$
\STATE \textbf{Stage 1: BO for Best Hyperparameters}
\STATE Use Algorithm 1 to select the optimal hyperparameters
\STATE \textbf{Return}: Optimal pre-trained model and hyperparameters
\STATE \textbf{Stage 2: Self-Adaptive Mechanisms}
\STATE Based on the optimal pre-trained model from Stage 1, perform the Algorithm 3  
\FOR{0:1000:5000 iterations}
    \STATE Perform ADAM optimization on the optimal pre-trained model
    \STATE Use Algorithm 2 to update the loss function weights
\ENDFOR
\STATE \textbf{Return}: New SA-model and sampling points distribution
\STATE \textbf{Stage 3: Training PINN}
\STATE Train the new SA-model using L-BFGS for 5000-10000 iterations
\STATE \textbf{Return}: Prediction solution$\hat{u}(x,t;\theta)$
\end{algorithmic}
\label{whole process}
\end{algorithm}

\begin{figure}[t]
\centering
\includegraphics[width=\linewidth]{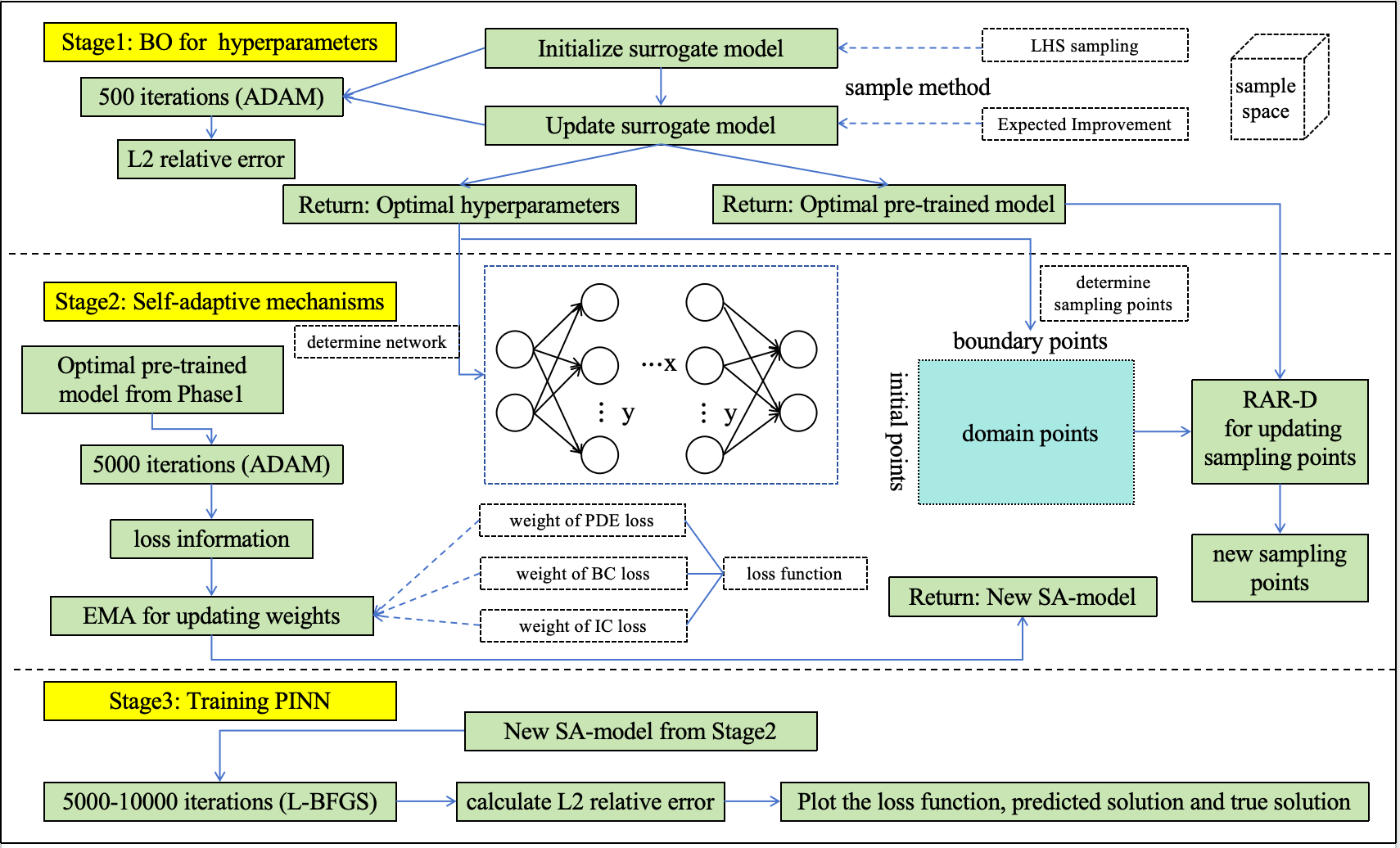}
\caption{Entire process of BO-SA-PINNs; Stage 1 is selecting suitable hyperparameters, stage 2 is using self-adaptive mechanisms to adjust loss function weights and sampling points distribution, stage 3 is stable optimization.}\label{process of BO-SA-PINNs}
\end{figure}

The overall process framework is shown in Figure \ref{process of BO-SA-PINNs}.

\section{Numerical examples}
\label{result}
In this section, numerical experiments will be conducted to demonstrate the performance of BO-SA-PINNs on various benchmarks. BO-SA-PINNs not only achieve higher accuracy and solution efficiency compared to the baseline PINNs, but also outperform improved PINNs like SA-PINNs in terms of both accuracy and solution efficiency in some test cases. To improve the efficiency of BO, we vary discrete hyperparameters using different step sizes. The number of neural network layers and hidden neurons increase by a step size of 1, while sampling points change in units of 50. For continuous hyperparameters, the loss function weights are adjusted with a step size of 0.0001, and the learning rate varies with a step size of 0.001. EMA decay parameter \( \beta \in (0, 1) \) and \( \gamma \in (0, 1) \) are both set to 0.999.

All experiments are conducted on an Nvidia RTX 4090D, with evaluation metrics being L2 relative error and the number of iterations. In addition, we also need to compare the number of sampling points and the complexity of the neural network of different methods, because the number of sampling points and the complexity of the neural network affect the computational cost of each iteration. We conduct 5 random experiments for each case and record the optimal L2 relative error. Our code is based on DeepXde \cite{lu2021deepxde} and can be obtained on \href{https://github.com/zr777777777777/BO-SA-PINNs.git}{our github}\footnote{The code is available at \url{https://github.com/zr777777777777/BO-SA-PINNs.git}}. We also present the optimal relevant hyperparameters for each numerical experiment in \ref{2}.

\subsection{2D Helmholtz equation}
The Helmholtz equation is typically used to describe the behavior of wave and diffusion processes, and can be employed to model evolution in a spatial domain or combined spatial-temporal domain. 
Here we study a particular Helmholtz PDE existing only in the spatial $(x,y)$ domain, described as:
\begin{equation}
u_{xx} + u_{yy} + k^2 \, u - s(x,y) = 0 
\end{equation}
\begin{equation}
u(-1, y) = u(1, y) = u(x, -1) = u(x, 1) = 0
\end{equation}
where $x \in [-1, 1]$, $y \in [-1, 1]$, $u(x,y)$ denotes the field variable and the forcing term is defined as:
\begin{eqnarray}
 s(x,y) &=& -(a_1 \pi)^2 \,\sin(a_1 \pi x)\,\sin(a_2 \pi y)
  - (a_2 \pi)^2 \,\sin(a_1 \pi x)\,\sin(a_2 \pi y) \nonumber\\[0.5ex]
      &&\quad + k^2 \,\sin(a_1 \pi x)\,\sin(a_2 \pi y)
\end{eqnarray}
The source term results in a closed-form analytical solution:
\begin{equation}
u(x,y) = \sin(a_1 \pi x)\,\sin(a_2 \pi y).
\end{equation}

\subsubsection{Comparative experiments and ablation experiments}
In this experiment, the total number of iterations is 15,500 (500, 5000, 10000 iterations for three stages respectively). We conducted comparative experiments with the baseline PINN and SA-PINN, whose experimental results are taken from \cite{wang2021understanding} and \cite{braga2021self}. We found that the L2 relative error of BO-SA-PINN was $3.21 \times 10^{-4}$, which is much smaller than the $1.40 \times 10^{-1}$ of the baseline PINN and the $3.20 \times 10^{-3}$ of SA-PINN, demonstrating the effectiveness of our method. In addition, we also performed ablation experiments on this problem to validate the importance of each improvement and the TG activation function has the greatest impact. Removing the TG activation function causes the performance to drop by about 3.7 times; after removing the EMA or SA mechanism, the performance drops by about 1.3 times and 1.4 times.The experimental results are shown in Table \ref{Helmholtz}.

\begin{table}[ht]
\caption{Comparison of L2 Error for different methods and ablation experiment results of BO-SA-PINN; No TG means that the TG activation function is not used, and $tanh(x)$ is used; No EMA means that there is no adjustment of the loss function weights; No SA mechanism means that there is no self-adaptive adjustment in stage2.}
\centering
\begin{tabular}{cccc}
\toprule
\textbf{Category} & \textbf{Method} & \textbf{L2 relative error} & \textbf{Iterations} \\
\midrule
\multirow{2}{*}{Baseline} & PINN & $1.40 \times 10^{-1}$ & 40000 \\
                          & SA-PINN & $3.20 \times 10^{-3}$ & 20000 \\
\midrule
\multirow{4}{*}{BO-SA-PINN} & BO-SA-PINN & $\mathbf{3.21 \times 10^{-4}}$ & \textbf{15500}\\
                            & No TG & $1.52 \times 10^{-3}$ & 15500\\
                            & No EMA & $7.34 \times 10^{-4}$ & 15500\\
                            & No SA mechanism & $7.75 \times 10^{-4}$ &15500\\
\bottomrule
\end{tabular}
\label{Helmholtz}
\end{table}

Regarding the number of sampling points, BO-SA-PINN uses 3,000 initial sampling points and 500 sampling points selected by RAR-D, while SA-PINN uses 100,400. Therefore, BO-SA-PINN uses approximately 96.5\% fewer sampling points compared to SA-PINN and reduces the cost of each iteration. Another interesting phenomenon in this case is that SA-PINN selects a large number of domain sampling points, while BO-SA-PINN selects more boundary sampling points. From the perspective of the cost of neural network training, the network architecture of BO-SA-PINN is [2,74,74,1], with a total number of parameters of 5847, and the network architecture of SA-PINN is [2,50,50,50,50,1], with a total number of parameters of 7851. The number of neural network parameters of BO-SA-PINN is significantly lower, 25.5\% less. In this case, the total training time is $614.5$ seconds. From this we can see that BO-SA-PINN can indeed improve training efficiency, both in terms of memory overhead and training time.

\subsubsection{Comparative experiments to verify the advantages of TG activation function}
\begin{figure}[t]
\centering
\includegraphics[width=\linewidth]{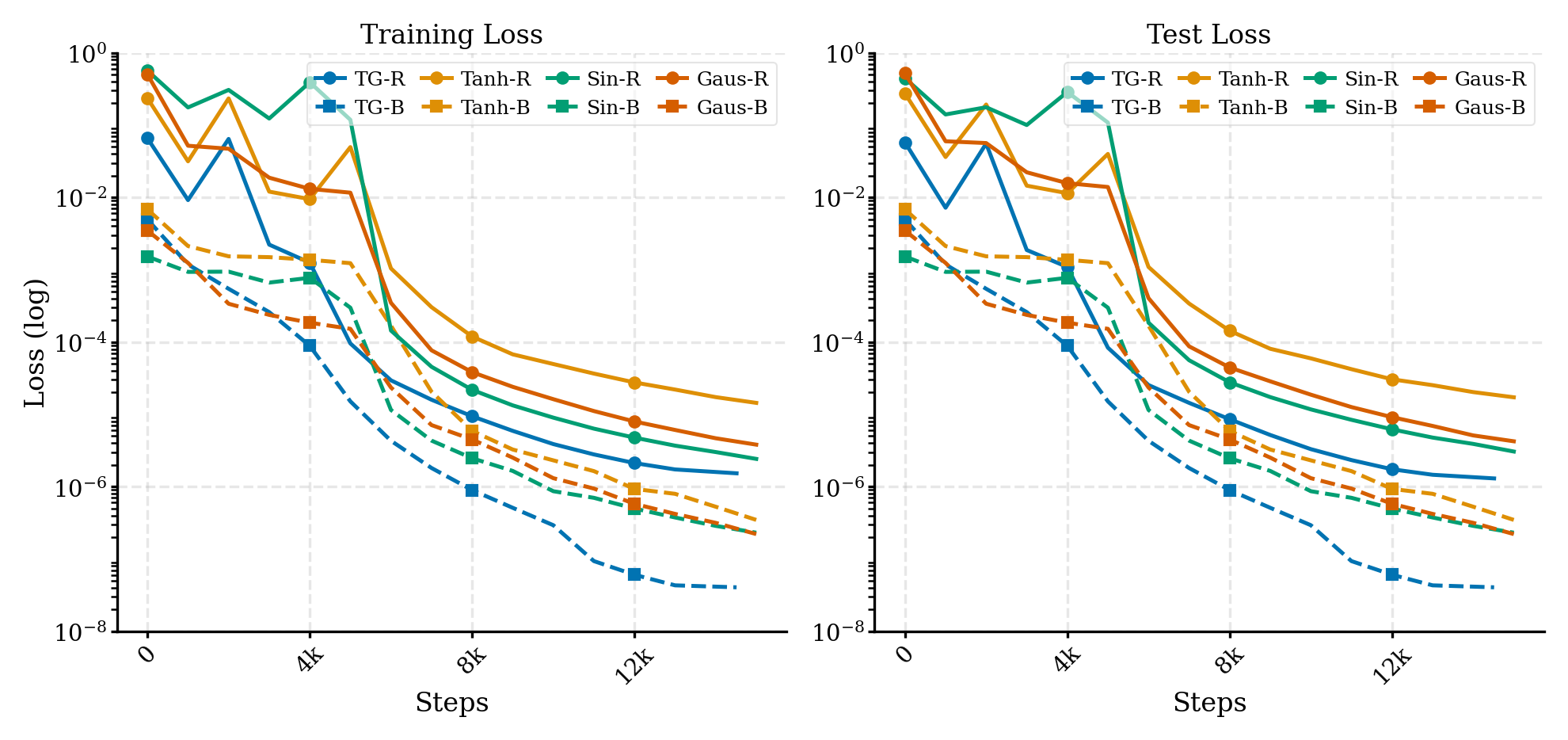}
\caption{Train loss and test loss of TG activation function (TG), $tanh(x)$ (Tanh), $sin(x)$ (Sin) and $e^{-\frac{x^2}{2}}$ (Gaus); R denotes residual loss and B denotes boundary loss}\label{TG_loss}
\end{figure}

In order to more intuitively demonstrate the optimization of the training process brought by the TG activation function, we show the loss function value during the training process. We can clearly see  that the training loss and test loss of TG activation function are both smaller than those of $tanh(x)$,$sin(x)$ and $e^{-\frac{x^2}{2}}
$ during training process in Figure \ref{TG_loss}, which indicates that the TG activation function has better optimization effect and generalization ability. Judging from the final L2 relative error in Table \ref{activation}, the TG activation function is better than $tanh(x)$, $sin(x)$ and $e^{-\frac{x^2}{2}}$ in this case.

\begin{table}[ht]
\caption{Comparison of L2 Error for different activation functions of BO-SA-PINN}
\centering
\begin{tabular}{cc}
\hline
\textbf{Activation} & \textbf{L2 relative error} \\
\hline
TG & $\mathbf{3.21 \times 10^{-4}}$   \\
$tanh(x)$ & $1.52 \times 10^{-3}$  \\
$sin(x)$ & $6.91 \times 10^{-4} $\\
$e^{-\frac{x^2}{2}}$ & $8.18 \times 10^{-4} $ \\
\hline
\end{tabular}
\label{activation}
\end{table}

\subsection{2D Maxwell equations for disk scattering— a complex number solver}
The 3D Maxwell equations can be simplified by making certain assumptions \cite{jin2015finite}. We consider a model that is infinitely extended along the $z$-direction, with a constant cross-sectional shape and dimensions throughout. Furthermore, we assume a uniformly incident wave along the $x$-direction, causing all partial derivatives with respect to $z$ to vanish. Our study is conducted in accordance with the $TM_z$ mode in frequency domain mode, which is described by the equations below:
\begin{equation}
\left\{\begin{aligned}
&\frac{\partial^2 E_{rz}}{\partial x^2}+\frac{\partial^2 E_{rz}}{\partial y^2}+w_r^2\mu\varepsilon_rE_{rz}=0 \\
&\frac{\partial^2 E_{iz}}{\partial x^2}+\frac{\partial^2 E_{iz}}{\partial y^2}+w_r^2\mu\varepsilon_rE_{iz}=0
\end{aligned}\right.
\end{equation}
where \( E_{rz} \) and \( E_{i\!z} \) denote the real and imaginary parts of the electric field intensity \( E \) in the \( z \)-direction, respectively,  $\varepsilon_r$ and \( \mu \) are the permittivity and permeability of the material.

In this problem, we solve Maxwell equations for \( E_{rz} \) and \( E_{iz} \) in a square region with a dielectric disk of different materials. The square geometry is assigned the electric and magnetic properties of vacuum. A monochromatic plane wave \((E_{inc}, H_{inc})\) is incident in the vacuum, propagating parallel to the \( z \)-axis. The considered geometry is a square with a side length of 2, on which a dielectric disk with a radius \( R = 0.25 \) and a relative permittivity of \( \varepsilon_c =1.0, 1.5, 4.0 \) is placed at the center. The addition of this disk introduces a possible discontinuity in the permittivity.

We consider absorbing boundary conditions, which are characterized by absorbing all electromagnetic waves at the boundary, making the problem in the truncated region equivalent to the original open-domain problem:
\begin{equation}
\left\{\begin{aligned}
\mathbf{n}\times\mathbf{E}_r - \sqrt{\frac{\mu}{\epsilon}}\mathbf{n}\times(\mathbf{H}_r\times\mathbf{n})&=\mathbf{n}\times\mathbf{E}_{r,inc}-\sqrt{\frac{\mu}{\epsilon}}\mathbf{n}\times(\mathbf{H}_{r,inc}\times\mathbf{n})\\
\mathbf{n}\times\mathbf{E}_i - \sqrt{\frac{\mu}{\epsilon}}\mathbf{n}\times(\mathbf{H}_i\times\mathbf{n})&=\mathbf{n}\times\mathbf{E}_{i,inc}-\sqrt{\frac{\mu}{\epsilon}}\mathbf{n}\times(\mathbf{H}_{i,inc}\times\mathbf{n})
\end{aligned}\right.,
\end{equation}
where, $\mathbf{n}$ is the unit outward normal vector on the truncated boundary, $\mathbf{H}_r$ and $\mathbf{H}_i$ represent the real and imaginary parts of the magnetic field $\mathbf{H}$, $\mathbf{E}^{inc}$ and $\mathbf{H}^{inc}$ represent the incident electric and magnetic fields, respectively. 

The analytical solution is:
\begin{equation}
\left\{\begin{aligned}
u(r,\theta)&=\sum_{n=-\infty}^{+\infty}i^n\left[J_n(Kr)+\alpha_nH_n^{(1)}(Kr)\right]e^{in\theta}\quad\text{for }r\leq R\\
u(r,\theta)&=\sum_{n=-\infty}^{+\infty}i^n\beta_nJ_n(Kr)e^{in\theta}\quad\text{for }r\geq R
\end{aligned}\right.
\end{equation}
where, $\alpha_n$ and $\beta_n$ have the following forms:
\begin{equation}
\alpha_n=\frac{\mu K_iJ_n'(K_iR)J_n(KR)-KJ_n(K_iR)J_n'(KR)}{KH_n^{(1)'}(KR)J_n(K_iR)-\mu K_iJ_n'(K_iR)H_n^{(1)}(KR)},
\end{equation}
\begin{equation}
\beta_n=\frac{KH_n^{(1)'}(KR)J_n(KR)-KH_n^{(1)}(KR)J_n'(KR)}{KH_n^{(1)'}(KR)J_n(K_iR)-\mu K_iJ_n'(K_iR)H_n^{(1)}(KR)},
\end{equation}
where $R$ is the radius of the disk, $H_n^{(1)}$ and $H_n^{(1)'}$ are the first-kind Hankel function and its derivative, $J_n$ and $J_n'$ are the first-kind Bessel function and its derivative.

\subsubsection{Comparative experiments and ablation experiments}

Possible factors affecting the solution accuracy in this problem include high frequency, discontinuous derivatives caused by non-homogeneous media, and the complex numbers. Therefore, the baseline PINN struggles to solve the problem accurately. The baseline PINN was tested with multiple experiments, with the hidden layers set to 50$\times$3, the learning rate of the ADAM optimizer set to 0.01, 5,000 sampling points within the region, and 2,500 boundary sampling points. Additionally, the residual loss and boundary loss weights were set to 0.0125 and 0.15, respectively (This is a set of hyperparameters that others have chosen to solve this problem.). 

We selected a frequency of 300 MHz, with the dielectric constant in the disk region set to 1.0, 1.5 or 4.0, and the dielectric constant in the square region set to 1.0. The real and imaginary parts of $E_z$ were solved under high-frequency and different dielectric contrast conditions to test the performance of BO-SA-PINN. We performed comparative experiments with baseline PINN, and we found that when the dielectric constant in the disk region was 1.0 which is not a scattering problem, the L2 relative error of BO-SA-PINN are $7.98 \times 10^{-5}$ and $1.11 \times 10^{-4}$, respectively, which are smaller than the baseline PINN's $4.66 \times 10^{-4}$ and $4.38 \times 10^{-4}$. When the dielectric constant in the disk region is 1.5 and 4.0, the L2 relative error of BO-SA-PINN are $1.74 \times 10^{-2}$ and $1.17 \times 10^{-2}$, $2.84 \times 10^{-2}$ and $6.28 \times 10^{-2}$, respectively, which are also smaller than the baseline PINN's L2 relative error. We also conducted ablation experiments in this case to further verify the positive role of each improvement and the complete results are recorded in Table \ref{Maxwell}.

\begin{table}[h]
\caption{L2 Relative Error of PINN, SA-PINN, BO-SA-PINN for 2D Maxwell equation and ablation experiment results}
\centering
\begin{tabular}{cccccc}
\toprule
\multirow{2}{*}{\textbf{Parameters}} & \multirow{2}{*}{\textbf{Method}} & \multicolumn{2}{c}{\textbf{L2 relative error}}& \multirow{2}{*}{\textbf{Iterations}} \\
\cline{3-4} 
& & \textbf{Real Part} & \textbf{Imaginary Part} \\
\midrule
\multirow{3}{*}{$\varepsilon_s=1, \varepsilon_c=1.0$} 
  & PINN & $4.66 \times 10^{-4}$ & $4.38 \times 10^{-4}$ &20000\\
  & BO-SA-PINN & $\mathbf{7.98 \times 10^{-5}}$ & $\mathbf{1.11 \times 10^{-4}}$ &10500 \\
  & No TG & $3.09 \times 10^{-4}$ & $2.66 \times 10^{-4}$&10500\\
\midrule
\multirow{5}{*}{$\varepsilon_s=1, \varepsilon_c=1.5$} 
  & PINN & $2.06 \times 10^{-2}$ & $2.30 \times 10^{-2}$ &20000 \\
  & BO-SA-PINN & $\mathbf{1.74 \times 10^{-2}}$ & $\mathbf{1.17 \times 10^{-2}}$ &15500\\
  & No TG & $1.73 \times 10^{-2}$ & $1.21 \times 10^{-2}$&15500 \\
  & No EMA & $1.85 \times 10^{-2}$ & $1.24 \times 10^{-2}$ &15500\\
  & No SA Mechanism & $2.03 \times 10^{-2}$ & $1.58 \times 10^{-2}$ &15500\\
\midrule
\multirow{5}{*}{$\varepsilon_s=1, \varepsilon_c=4$} 
  & PINN & $2.07 \times 10^{-1}$ & $1.59 \times 10^{-1}$ &20000\\
  & BO-SA-PINN & $\mathbf{2.84 \times 10^{-2}}$ & $\mathbf{6.28 \times 10^{-2}}$ &15500\\
  & No TG & $2.88 \times 10^{-2}$ & $6.32 \times 10^{-2}$ &15500\\
  & No EMA & $2.94 \times 10^{-2}$ & $6.68 \times 10^{-2}$ &15500\\
  & No SA Mechanism & $3.91 \times 10^{-2}$ & $8.03 \times 10^{-2}$ &15500\\
\bottomrule
\end{tabular}
\label{Maxwell}
\end{table}

\begin{figure}[t]
\centering
\includegraphics[width=\linewidth]{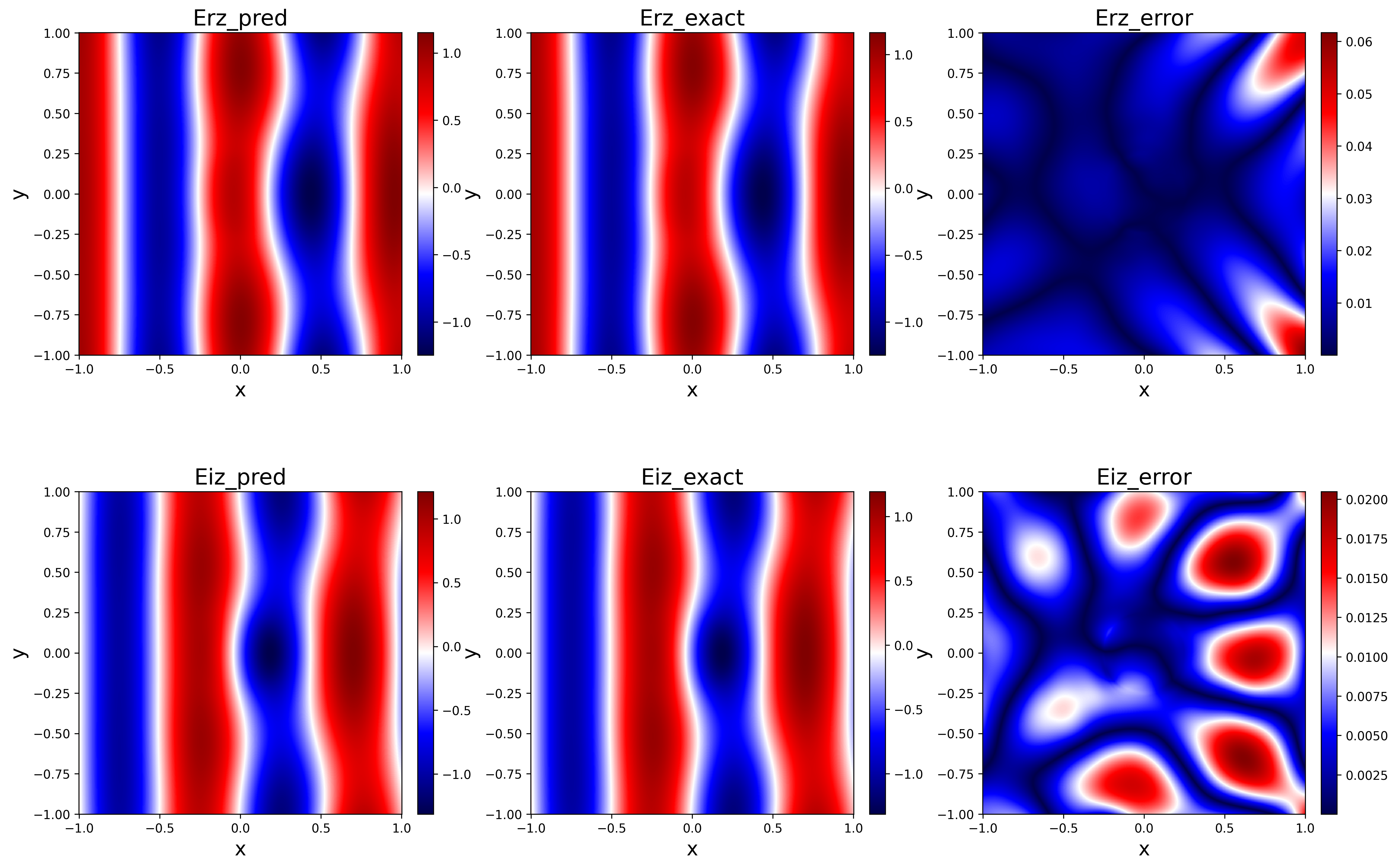}
\caption{Predicted solution, true solution, and error of 2D Maxwell problem of BO-SA-PINN in $f=300$MHz, $\varepsilon_s=1, \varepsilon_c=1.5$, including real part and imaginary part}\label{fig5}
\end{figure}

\subsubsection{Experiments to verify the effectiveness of BO-SA-PINNs on ResNet}

In order to verify that this framework is not only effective for the hyperparameter selection of FCNN, but also has similar effects on other network types, we chose to use ResNet for verification. The problem parameters are set to $f=300$MHz, $\varepsilon_s=1, \varepsilon_c=4$. The experimental results are shown in Table \ref{ResNet}. From the experimental results, we can see that the BO-SA-PINN framework can be applied to ResNet, and ResNet may be able to achieve an approximately accurate solution faster than FCNN. Thus, in complex problem situations, perhaps ResNet would be a better choice.

\begin{table}[h]
\caption{L2 Relative Error of different network types for 2D Maxwell equation}
\centering
\begin{tabular}{cccccc}
\toprule
\multirow{2}{*}{\textbf{Network type}} & \multirow{2}{*}{\textbf{Method}} & \multicolumn{2}{c}{\textbf{L2 relative error}} & \multirow{2}{*}{\textbf{Iterations}} \\
\cline{3-4} 
& & \textbf{Real Part} & \textbf{Imaginary Part} & \\
\midrule
\multirow{2}{*}{FCNN} 
  & PINN & $2.07 \times 10^{-1}$ & $1.59 \times 10^{-2}$ & 20000 \\
  & BO-SA-PINN &$2.84 \times 10^{-2}$ & $6.28 \times 10^{-2}$ & 15500 \\
\midrule
\multirow{2}{*}{ResNet} 
  & PINN & $4.24 \times 10^{-2}$ & $8.16 \times 10^{-2}$ & 15000 \\
  & BO-SA-PINN & $3.00 \times 10^{-2}$ & $6.29 \times 10^{-2}$ & 8500 \\
\bottomrule
\end{tabular}
\label{ResNet}
\end{table}

\subsection{Viscous Burgers equation}
We define a Burgers equation:
\begin{equation}
\frac{\partial u}{\partial t} + u \frac{\partial u}{\partial x} = \nu \frac{\partial^2 u}{\partial x^2}, \quad x \in [-1, 1], \quad t \in [0, 1]
\end{equation}

with the Dirichlet boundary conditions and initial conditions:
\begin{equation}
u(-1,t) = u(1,t) = 0, \quad u(x,0) = -\sin(\pi x).
\end{equation}

The inputs of 1D viscous Burgers equation are \( x \) and \( t \), and the output is \( \hat{u}(x,t;\theta) \). The reference solution data for this problem comes from DeepXDE \cite{lu2021deepxde}. To solve this problem, the required number of iterations was 5500 (with 500, 2000, and 3000 iterations in three stages, respectively). The final output L2 relative error is $\mathbf{3.56 \times 10^{-4}}$, and if we choose to use $tanh(x)$, the L2 relative error is \( 3.59 \times 10^{-4} \), which are both better than the \( 4.80 \times 10^{-4} \) from SA-PINN\cite{braga2021self} and the \( 6.70 \times 10^{-4} \) from the baseline PINN\cite{raissi2019physics}. In this case, BO-SA-PINN uses 7250 sampling points and 500 sampling points selected by RAR-D, while SA-PINN uses 10400 sampling points, thus using 25.5\% fewer sampling points. However, the number of parameters of the neural network will be larger in this case.

\subsection{nD Poisson equation}
Compared to traditional numerical methods, PINN has the significant advantage of solving high-dimensional problems. Therefore, we validate the performance of BO-SA-PINN on the 3D, 4D, and 5D Poisson equation and compare it with the baseline PINN. We are studying the following n-dimensional Poisson equation defined on the spatial domain \(x \in \Omega = [-1, 1]^n\):
\begin{equation}
\Delta u(x) = \sum_{i=1}^{n} \exp(x_i)
\end{equation}
\begin{equation}
u(x) = 0 \quad \text{on} \quad \partial \Omega
\end{equation}

This source term leads to a closed-form analytical solution for the Poisson equation, which is:
\begin{equation}
u(x) = \sum_{i=1}^{n} \exp(x_i)
\end{equation}

The total number of iterations is set to 5500. To solve this problem, the required number of iterations was 5500 (with 500, 2000, and 3000 iterations in three stages, respectively). Since we do not know the hyperparameters chosen by others, we select the hyperparameters of the baseline PINN according to our own experience in this experiment, which is also in line with the actual experimental scenario. For the baseline PINN, there are 2500 ADAM iterations and 3000 L-BFGS iterations. For the baseline PINN, we use a 50×3 hidden layer, set the learning rate of the ADAM optimizer to 0.001, use 5000 interior sampling points, and 3500 boundary sampling points. Additionally, both the residual loss and boundary loss weights are set to 1. The experiment results are shown in Table \ref{Poisson}.

\begin{table}[ht]
\caption{Comparison of BO-SA-PINN and PINN for different dimensions Poisson equation}
\centering
\begin{tabular}{ccc}
\toprule
\textbf{Dimension} & \textbf{Method} & \textbf{L2 relative error} \\
\midrule
\multirow{2}{*}{3} & BO-SA-PINN & $4.52 \times 10^{-5}$ \\
                  & PINN & $3.13 \times 10^{-1}$ \\
\midrule
\multirow{2}{*}{4} & BO-SA-PINN & $6.57 \times 10^{-5}$ \\
                  & PINN & $2.48 \times 10^{-1}$ \\
\midrule
\multirow{2}{*}{5} & BO-SA-PINN & $9.00 \times 10^{-5}$ \\
                  & PINN & $2.52 \times 10^{-1}$ \\
\bottomrule
\end{tabular}
\label{Poisson}
\end{table}

\section{Conclusion}
\label{conclusion}
In this paper, we propose BO-SA-PINNs, a novel multi-stage PINN framework that can automatically design suitable PDE solvers and adaptively optimize the sampling point distribution and loss function weights based on training information, which can greedily maximize the performance of the PDE solvers. And a new activation function TG is proposed and has been illustrated to be effective. Comparative experiments for solving the 2D Helmholtz, 2D Maxwell and 1D Burgers equations show that BO-SA-PINNs outperform baseline PINNs and perform better than SA-PINNs in some cases. The ablation experiments verify the positive effect of each improvement including the TG activation function and self-adaptive mechanisms. Experimental results for solving the nD Poisson equation demonstrate that BO-SA-PINNs have the potential for efficiently solving high-dimensional problems. In total, BO-SA-PINNs can achieve higher accuracy and efficiency in many cases.

We believe that BO-SA-PINNs open up new possibilities for the application of deep neural networks in both forward and inverse modeling in engineering and science. However, there are still many improvements needed in PINNs. For example, since using the TG activation function in this problem will cause possible overfitting in the third stage in some extreme cases but the TG activation function performs much better than $tanh(x)$ in the early stage of training, we hope to find a way to prevent the overfitting problem in the future study. We can also develop optimization algorithms better suited for PINNs. Furthermore, ensuring that BO can find the global optimum and improving the generalization ability of PINNs through uncertainty metrics are also meaningful research directions.

\appendix
\section{Justification of TG activation function}
\label{appA}

Universal Approximation Theorem states that a feedforward neural network with a single hidden layer and sufficient hidden units can approximate any continuous function to an arbitrary degree of precision, given that the activation function satisfies certain conditions. For the TG activation function $\phi(x) = \tanh(x)e^{-x^2/2}$, it satisfies the conditions of the theorem, and the specific proof is as follows:

$\phi(x) \text{ is the product of two continuous functions } \tanh(x) \text{ and } e^{-x^2/2}.$
Since the product of continuous functions is also continuous, $\phi(x)$ is continuous on $\mathbb{R}$.

Suppose $\phi(x)$ is a polynomial function, then its Taylor expansion should have a finite number of terms. However:
\[
\tanh(x) = x - \frac{x^3}{3} + \frac{2x^5}{15} - \dots \text{an infinite series}
\]
\[
e^{-x^2/2} = 1 - \frac{x^2}{2} + \frac{x^4}{8} - \dots \text{an infinite series}
\]
Thus, $\phi(x)$, has an infinite number of terms in its Taylor expansion, and as $|x| \to \infty$, $\phi(x) \to 0$. Any non-zero polynomial, however, will not tend to zero as $x$ goes to infinity. Therefore, $\phi(x)$ cannot be a polynomial function.

According to the conclusion that any continuous and non-polynomial activation function satisfies the conditions for the Universal Approximation Theorem\cite{leshno1993multilayer}, TG is qualified to be an activation function. Specifically, for any continuous function $f: K \to \mathbb{R}$ defined on a compact set $K \subset \mathbb{R}^n$, there exists a single hidden-layer neural network that can approximate $f$ to any desired degree of precision by adjusting weights and biases.

\section{Specific selecting hyperparameters in numerical experiments}
\label{2}

We share the optimal hyperparameters selected in numerical experiments. If you want to solve similar PDEs, you may choose them directly from the Table B.7.

\begin{table}[ht]
\caption{Selected hyperparameters in numerical experiments}
\centering
\begin{tabular}{lccccccccc}
\toprule
PDEs & \(\omega_R\) & \(\omega_B\) & \(\omega_I\) & \(lr\) & \(n_d\) & \(n_b\) & \(n_i\) & layer & neurons \\
\midrule
2D Helmholtz       & 0.0144 & 0.2406 & - & 0.005 & 1350  & 1650  & - & 2 & 74  \\
2D Maxwell(1.0)    & 0.0848 & 0.1215 & - & 0.007 & 1550  & 1200  & - & 6 & 48  \\
2D Maxwell(1.5)    & 0.0225 & 0.1081 & - & 0.008 & 3150  & 2500  & - & 5 & 37  \\
2D Maxwell(4.0)    & 0.0327 & 0.1810 & - & 0.007 & 4700  & 1500  & - & 5 & 21  \\
1D Burgers         & 0.0410 & 0.0593 & 0.0815 & 0.006 & 4250  & 1900  & 1100  & 7 & 35  \\
3D Poisson         & 0.0767 & 0.1142 & - & 0.007 & 13600 & 1050  & - & 3 & 74  \\
4D Poisson         & 0.0328 & 0.2335 & - & 0.009 & 12150 & 6500 & - & 6 & 59  \\
5D Poisson         & 0.0377 & 0.1565 & - & 0.009 & 9550  & 2700  & - & 3 & 58  \\
\bottomrule
\end{tabular}
\end{table}

\bibliographystyle{elsarticle-num}
\bibliography{reference}

\end{document}